\documentclass[prd,twocolumn]{revtex4-1}

\usepackage{amsmath, amssymb}
\usepackage[]{graphicx}
\usepackage[caption=false]{subfig}
\usepackage{hyperref}

\begin{document}

\title{Bounds on QCD axion mass and primordial magnetic field from CMB $\mu$-distortion}
\author{Damian Ejlli}
\email{damian.ejlli@lngs.infn.it}

\affiliation{Theory group, INFN Laboratori Nazionali del Gran Sasso, 67100 Assergi, Italy}
\affiliation{Department of Physics, Novosibirsk State University, Novosibirsk 630090 Russia}

\date{\today}
\begin{abstract}
The oscillation of the CMB photons into axions can cause CMB spectral distortion in the presence of large scale magnetic field. With the COBE limit on the $\mu$ parameter and a homogeneous magnetic field with strength $B\lesssim 3.2$ nG at the horizon scale, an upper limit on the axion mass is found to be, $m_a\lesssim 4.8\times 10^{-5}$ eV for the KSVZ axion model. On the other hand, using the value of excluded axion mass $m_a\simeq 3.5\times 10^{-6}$ eV from the ADMX experiment together with the COBE bound on $\mu$, is found $B\simeq 46$ nG for the KSVZ axion model and $B\simeq 130$ nG for DFSZ axion model, for a homogeneous magnetic field with coherence length at the present epoch $\lambda_B\simeq 1.3$ Mpc. Forecast on $B$ and $m_a$ for PIXIE/PRISM expected sensitivity on $\mu$ are derived. If CMB $\mu$ distortion would be detected by the future space missions PIXIE/PRISM and assuming that the strength of the large scale magnetic field is close to its canonical value, $B\simeq 1-3$ nG, axions in the mass range $2\, \mu$eV - $3\, \mu$eV would be potential candidates of CMB $\mu$-distortion.
\end{abstract}

\maketitle

The cosmic microwave background (CMB) presents small temperature anisotropy of the order of $\delta T/T\sim 10^{-5}$ on small angular scale and its spectrum is supposed to be slightly distorted  \cite{Zeldovich:1969ff} due to various mechanisms which might have operated in the early universe. In general these distortions are described in the terms of the so-called $\mu, i$ and $y$ parameters which their values quantify the type of each distortion \cite{chluba}. COBE \cite{Fixsen:1996nj} space mission obtained stringent limits on $|\mu|<9\times 10^{-5}$ and $|y|<1.5\times 10^{-5}$ parameters, thus implying that there might be a very narrow window to look for process leading to spectral distortion. Other planned space missions include PIXIE \cite{Kogut:2011xw} and PRISM \cite{Andre:2013afa} which expect to reach better sensitivity on $\mu$ and $y$ with respect to COBE of the order of $\mu \simeq 5\times 10^{-8}$ and $y\simeq 10^{-8}$.

Generally speaking the most popular proposed mechanisms which can create spectral distortion, can be classified as ''secondary'' mechanisms in the sense that the original CMB spectrum is affected indirectly. Indeed, in these models energy and photon number are injected into the medium from external sources such as decaying dark matter particles \cite{Chluba:2013wsa}, sound waves \cite{Khatri:2011aj} etc. On the other hand, CMB can also have ''primary'' spectral distortions which can be disentangled from the secondary ones. An interesting mechanism which can be classified as primary, is oscillation of the CMB photons into light bosons such as axions, axionlike particles (ALPs) and gravitons. These processes, in cosmological context, are possible in the presence of an external magnetic field where the photon has a vertex coupling with them. In the case of axions the relevant term which describes coupling of photons with axions is given by the interaction Lagrangian density
\begin{equation}\label{axion-photon}
\mathcal L_{a\gamma}=-\frac{g_{a\gamma}}{4}a\,F_{\mu\nu}\tilde F^{\mu\nu},
\end{equation}
where $F_{\mu\nu}$ is the electromagnetic field tensor, $\tilde F^{\mu\nu}$ is its dual and $a$ is the axion field. In general, the coupling constant of axions can be written as
\begin{equation}\label{axion-coupling}
g_{a\gamma}=\frac{\alpha_s}{2\pi f_a}\, \left(\frac{E}{N}-\frac{2}{3}\frac{4+w}{1+w}\right),
\end{equation}
where $\alpha_s$ is the fine structure constant, $f_a$ is the axion decay constant, $E$ is the electromagnetic anomaly associated with axial current and $N$ is the color anomaly. Among of all axion models, two of them namely the KSVZ  \cite{Shifman:1979if} and DFSZ \cite{Dine:1981za} axion models have been extensively studied in the literature. For the KSVZ model we have $E/N=8/3$ and $E/N=0$ for the DFSZ model. In both models, the coupling constant of axions to photons $g_{a\gamma}$ is proportional related to axion mass $m_a$. The latter is related with quark masses up (u) and down (d) and the relation between axion mass $m_a$ and axion decay constant $f_a$ is given by
\begin{equation}\label{axion-mass}
m_a=\frac{m_\pi\,f_\pi}{f_a}\frac{w^{1/2}}{1+w},
\end{equation}
where $m_\pi=135$ MeV is the pion mass, $f_\pi\simeq 92$ MeV is the pion decay constant and $w=m_u/m_d$ with $m_u, m_d$ being respectively the up and down quark masses.  The range of the parameter $w$ is between $0.35\leq w\leq 0.6$ \cite{pdg} where in general its standard value is taken $w=0.56$. For recent reviews on axions and ALPs see Ref. \cite{Dias:2014osa} and for earlier works on axions in cosmology see Ref. \cite{Kholopov}

The origin of the large scale magnetic field (which makes possible transition of photons into axions), is interesting by itself since its presence, would have enormous impact in several situations in cosmology (such as bing bang nucleosynthesis, CMB temperature anisotropy etc.) and in astrophysics (such as cosmic rays deflection etc). Thus, its strength $B_e$ and its direction are of fundamental importance. The most common ways to constrain large scale magnetic field strength have been essentially from CMB temperature anisotropy and Faraday rotation of the CMB \cite{Kronberg:1993vk}. In the former case, it is supposed that the external magnetic field would contribute to the total energy density of the universe, and therefore it would be possible that this additional energy density could cause CMB temperature anisotropy \cite{zel'dovich70}. In the latter case, the presence of magnetic field would cause polarisation of the CMB, through the so-called Faraday effect, namely the rotation of the polarisation plane of the CMB. It has also been shown that the Faraday effect can be induced by a coupling of a quintessential background field with pseudo-scalar coupling to the CMB, see Ref. \cite{Giovannini:2004pf} (for a link between the Faraday effect and CMB B-mode polarisation see Ref.  \cite{Giovannini:2014wta}).  For a review on large scale magnetic fields see Ref. \cite{Grasso:2000wj}

In a previous work \cite{Ejlli:2013uda}, we obtained tight limits on the ALP parameter space by using coupling of CMB photons with ALPs in primordial magnetic field. In this letter we study oscillation of CMB photons into axions in presence of large magnetic field and derive new limits on axion mass and magnetic field strength. Photon-axion mixing is phenomenologically different from oscillation into ALPs, since in the axion case the two quantities which characterize axions, its mass $m_a$ and coupling constant to photons $g_{a\gamma}$, are directly proportional with each other.  Consequently, in the case of photon-axion mixing the number of independent parameters is reduced to only $B_e$ and $m_a$ or $g_{a\gamma}$ with respect to the photon-ALP mixing. Therefore based on phenomenological or experimental results it would be possible that known one of the parameters $B_e$  or $m_a$, we can constrain the remaining one. 

Firstly, knowing the upper bounds on the magnetic field strength at the present time, we can find limits for the axion mass. In this case case the field strength and coherence length are fixed \emph{a priori}. Secondly, if we know experimental limits on the axion mass we can bound the magnetic field strength and discuss about its coherence length \emph{a posteriori}. In this letter we consider only uniform (homogeneous) magnetic field. The effect on the CMB oscillation due to non homogeneous (stochastic) magnetic field will not be considered. In connection with the first case, we use limits on the magnetic field from the CMB temperature anisotropy and Faraday rotation, where the field coherence length is greater or comparable with horizon scale. For a magnetic field with coherence length comparable to the horizon scale, CMB temperature anisotropy gives $B\lesssim 4$ nG \cite{Barrow:1997mj} and Faraday rotation of Lyman-$\alpha$ forest gives \cite{Blasi:1999hu}, $B\lesssim 1$ nG. As far as for the second case, we consider existing limits on axion mass to constrain strength of the homogeneous magnetic field with coherence length at least comparable to the horizon scale during $\mu$ epoch. In the formalism of the density matrix which we use below, the magnetic field is assumed to be homogeneous at given coherence length $\lambda_B$, where the field strength changes only due to the expansion of the universe. Here we adopt the rationalized Lorentz-Heaviside natural units, $c=\hbar=k_B=\epsilon_0=\mu_0=1.$ 

The study of oscillation of the CMB photons into axions with an essential loss of coherence is best formulated in the terms of the density operator of the system $\hat\rho$ (in our case the system is composed of axions and photons). To the linear order of approximation, it satisfies the quantum kinetic equation \cite{Dolgov:2002wy}
\begin{equation}\label{dens-ope}
\frac{d\hat\rho}{dt}=-i[\hat H, \hat\rho]-\{\hat\Gamma, (\hat\rho-\hat\rho_{eq})\},
\end{equation}
where $\hat H$ is the Hamiltonian of photon-axion system including refraction index (first order effects), $\hat\Gamma$ is the coherence breaking operator of photons and axions with the background medium, and $\hat\rho_{eq}$ is the equilibrium density operator. Since the magnetic field mix only the ($\times$) photon state (see below) with the axion, the matrix elements of the operators $\hat\rho$, $\hat\Gamma$ and $\hat H$ in the basis spanned by the two component field $\Psi^T=(A_\times, a)$ are respectively given by
\begin{equation}
\rho=
\begin{pmatrix}
  n_\gamma & \rho_{\gamma a} \\
  \rho_{a\gamma}  &  n_{a} \\
   \end{pmatrix},\quad
   \Gamma=
\begin{pmatrix}
  \Gamma_\gamma & 0 \\
  0 &  \Gamma_{a} \\
   \end{pmatrix},\quad H=\begin{pmatrix}
  M_{\times} & M_{a\gamma} \\
M_{a\gamma} & M_a\\
   \end{pmatrix}\end{equation}
where $\rho_\gamma=n_\gamma, \rho_a=n_a$ are respectively the photon and axion occupation numbers, $\rho_{\gamma a}=\rho_{a\gamma}^*=R+iI$ with $R$ and $I$ being respectively the real and the imaginary part of $\rho_{\gamma a}$. The matrix elements of the equilibrium density operator in the flavor space, are given by the equilibrium occupation number $n_{eq}=1/(e^x-1)$ times the identity matrix $\mathbf I$, $\rho_{eq}=n_{eq}\,\mathbf I$ where $x=\omega/T$ with $T$ being the photon temperature. The coherence breaking matrix ($\Gamma$), is diagonal in the flavor space and its entries are respectively given by the \emph{sum} of the scattering and the annihilation/absorption rates of photons ($\Gamma_\gamma$) and axions ($\Gamma_a$). Matrix elements which enter the interaction Hamiltonian, are respectively \cite{Raffelt:1987im} $M_\times=\omega(n-1)_\times$, $M_{a\gamma}=g_{a\gamma}B_T/2$, $M_a=-m_a^2/2\omega$. Here $B_T$ is the strength of the external magnetic field $\mathbf B_e$,
which is transverse to the direction $\mathbf x$,  of the photon/axion propagation.  
$A_{+, \times}$ are the photon polarization states with ${+, \times}$ being the polarization indexes (helicity) of the photon. 
The helicity state $(+)$ corresponds  to the polarization perpendicular to the external magnetic field and $(\times)$ describes the polarization parallel to the external field. For the purpose of this work and the cosmological epoch which we are interested in, the total refraction index is given by the sum of two main components: the refraction index due to electronic plasma $n_\textrm{pla}$ and refraction index due to vacuum polarization $n_\textrm{QED}$. The refraction index due to electronic plasma is given by $(n_\textrm{pla}-1)_{\times, +}=-\omega_\textrm{pla}^2/2\omega^2$ where $\omega_\textrm{pla}^2=4\pi n_e/m_e$ with $n_e$ being the number density of free 
electrons in the plasma. 
The refraction index due to QED effects, for $\omega\ll (2m_e/3)\times (B_c/B)$, is given by Ref. \cite{Tsai:1974fa} $(n-1)_{\times, +} =(\alpha/4\pi)\left(B_T/B_c\right)^2\left[\left(14/45\right)_{\times}, \left(8/45\right)_{+}\right]$,
where $B_c=m_e^2/e=4.41\times 10^{13}$ G is the critical magnetic field.

When total interaction rate which enter the problem is much bigger than expansion rate $\Gamma\gg H$ and photon-axion oscillation frequency $\omega_{osc}\gg H$, equation of motion for density matrix are given by steady state approximation, see Ref. \cite{Dolgov:2002wy} for details. In this case it is possible to express the imaginary part $I$ and real part $R$ through $n_\gamma$ and $n_a$, see Ref. \cite{Ejlli:2013uda} for more details. Moreover, if the interaction rate of axions with the medium is small, we can approximate the interaction rate of axions with the medium in Eq. \eqref{dens-ope}, as $\Gamma_a\simeq 0$. Indeed, this is a good approximation for the cosmological epoch which we are interested in and for the axion mass range we are going to consider (see below).
Also assuming that the photon-axion transition is dominated by the resonance, one can find an analytic solution for the production probability of axions at the resonance temperature $\bar T$
\begin{equation}  \label{prob-axion-1}
P_a(\bar T)=-\left.\frac{2\pi M_{a\gamma}^2}{kHT}\right|_{T=\bar T},
\end{equation}
where $M_{a\gamma}(\bar T)=(g_{a\gamma} B_0/2)\left(\bar T/T_0\right)^2$ and $k(\bar T)=d(\Delta M)/dT|_{T=\bar T}$ with $\Delta M(\bar T)=M_\times(\bar T)-M_a(\bar T)=M_\textrm{QED}(\bar T)-M_\textrm{pla}(\bar T)-M_a(\bar T)$. Here $M_\textrm{QED}$ and $M_\textrm{pla}$ are respectively the QED and plasma contributions to the refraction index in $\Delta M$. The field strength of the transverse part of magnetic field, $B_T$, scales with temperature as $B_T\sim B=B_0(\bar T/T_0)^2$ (magnetic flux conservation) with $B_0$ being the strength of magnetic field at present epoch. The term $H(\bar T)\bar T$ can be written as $H(\bar T)\bar T=H_0T_0\sqrt{\Omega_R}(\bar T/T_0)^3$ where $\Omega_R=9.21\times 10^{-5}$ is the present day density parameter of relativistic particles (photons and nearly massless neutrinos). During the $\mu$ epoch, the universe is radiation dominated  where ionization fraction of free electrons is unity, $X_e=1$. In this case we can expand $k(T)$ up to first order in power series and write $k(\bar T)=(3/\bar T)\left[M_\textrm{QED}(\bar T)-M_\textrm{pla}(\bar T)\right]$. Inserting all necessary terms into Eq.  \eqref{prob-axion-1} we get the following expression for $P_a$ at the resonance temperature $\bar T$
\begin{equation}\label{Axion-proba}
P_a(\bar T)=-\frac{2\pi}{3H(\bar T)}\frac{M_{a\gamma}^2(\bar T)}{M_\textrm{QED}(\bar T)+M_a(\bar T)},
\end{equation}
where in deriving Eq. \eqref{Axion-proba} we have used the fact that for $T=\bar T$ we have $\Delta M(\bar T)=0$. We may note that in the case $M_\textrm{QED}(\bar T)=-M_a(\bar T)$, the denominator of Eq. \eqref{Axion-proba} is zero and the probability goes to infinity. In such case one must consider the expansion of $\Delta M(T)$ up to the second order in $T$ around the resonance temperature $\bar T$. However, for our purpose we do not need it here.

In order to confront Eq. \eqref{Axion-proba} with the numerical results and because is more easy to calculate, let us consider the case when $M_\textrm{QED}\ll M_a$. In the redshift of interest for $\mu$-distortion and the photon energy considered here, the QED term in $M_\times$ is small with respect to the plasma term and therefore from the resonance condition $\Delta M=M_\times-M_a=0$ we get 
\begin{equation}\label{res-temp}
\left(\frac{\bar T}{T_0}\right)=9\times 10^6\, n_e^{-1/3}\bar m_a^{2/3}\quad \textrm{cm}^{-1},
\end{equation}
where $\bar m_a=m_a/\textrm{eV}$, $n_e\simeq 0.88\, n_B(T_0)$ is the number density of the free electrons at the present epoch and $n_{B}(T_0)=2.47\times 10^{-7}$ cm$^{-3}$ is the number density of baryons. Eq. \eqref{res-temp} is a constraint relation for the axion mass in the resonant case. Inserting all necessary quantities into Eq. \eqref{Axion-proba} we get the following expression for $P_a$
\begin{equation}\label{res-probab}
P_a(\bar T)=5.75\times 10^{-27}\,x\,C_{a\gamma}^2\,B_\textrm{nG}^2 \left(\frac{\bar T}{T_0}\right)^3,
\end{equation}
where $B_\textrm{nG}=(B_0/\textrm{nG})$ and $C_{a\gamma}$ is defined as 
\begin{equation}
C_{a\gamma}\equiv  \left(\frac{E}{N}-\frac{2}{3}\frac{4+w}{1+w}\right)\frac{1+w}{w^{1/2}},
\end{equation}
where for $w=0.56$, $|C_{a\gamma}|\simeq 4$ for $E/N=0$ (KSVZ model) and $|C_{a\gamma}|\simeq 1.49$ for $E/N=8/3$ (DFSZ model). It is important to emphasize that Eq. \eqref{res-probab} is valid when $M_\textrm{QED}\ll M_a$ or 
\begin{equation}\label{B-limit}
B_\textrm{nG}^{1/3}\,x^{1/3}\left(\frac{T}{T_0}\right)\ll 1.23\times 10^9\,\bar m_a^{1/3}.
\end{equation}
On the other hand, we also need to calculate the axion mass at the resonance temperature $\bar T$ which is given by Eq. \eqref{res-temp}. Assuming that the interested temperature interval is coincident with the $\mu$ epoch, $2.88\times 10^5\lesssim T/T_0\lesssim 2\times 10^6$, the axion mass in this interval is
\begin{equation}\label{mass-range}
2.66\times 10^{-6}\, \textrm{eV}\,\lesssim \bar m_a\lesssim 4.88\times 10^{-5}\, \textrm{eV}.
\end{equation}
So, as far as we limit our consideration for the magnetic field strength of the order $B_\textrm{nG}\lesssim 10^3$ and axion mass range given by Eq. \eqref{mass-range}, we can safely use Eq. \eqref{res-probab}. 

\begin{figure*}[htbp!]
\centering
\mbox{
\subfloat[\label{fig:Fig1}]{\includegraphics[scale=0.6]{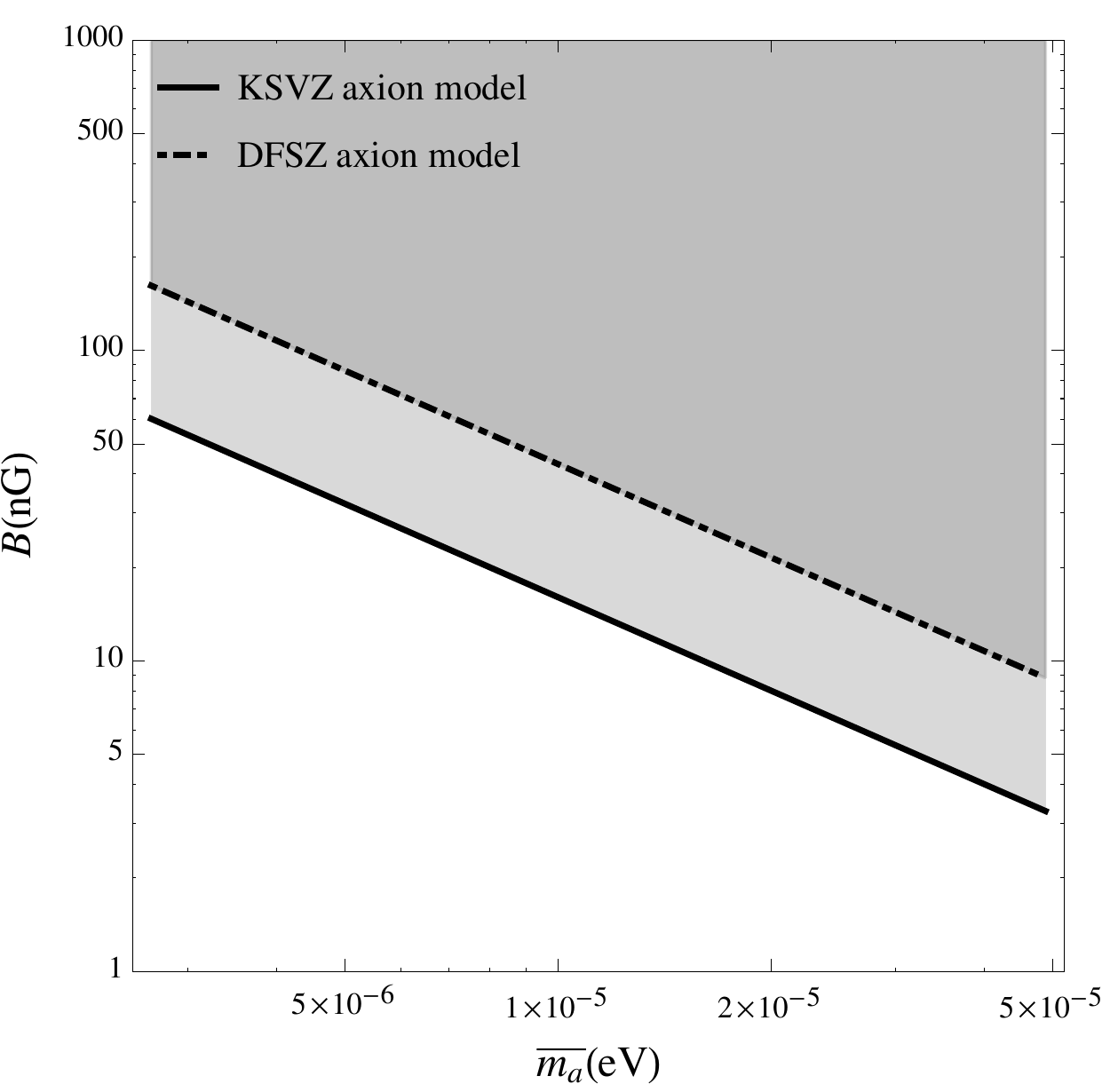}}\qquad
\subfloat[\label{fig:Fig2}]{\includegraphics[scale=0.6]{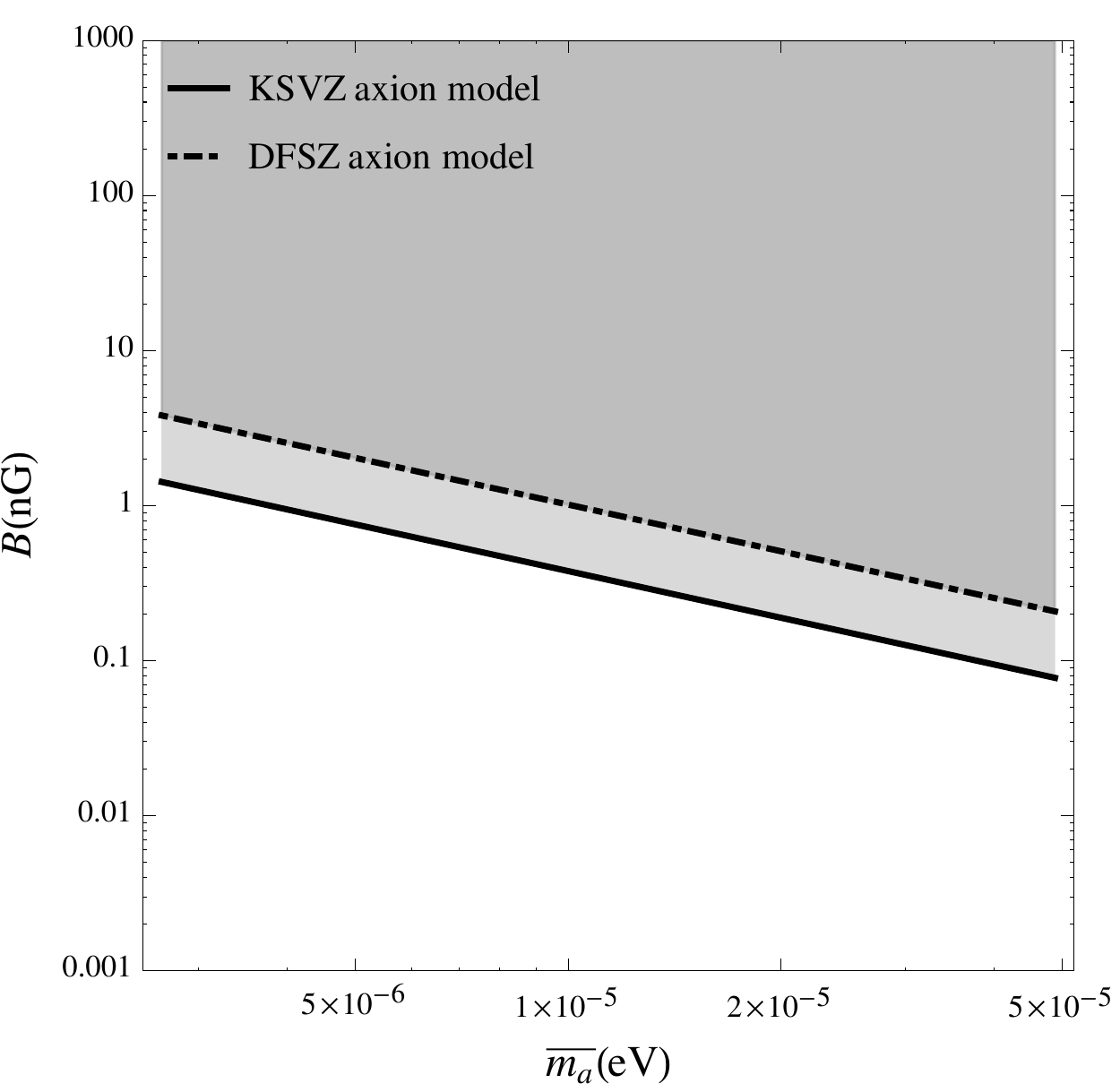}}}
\caption{Exclusion and sensitivity plot for the axion parameter space  $B-\bar m_a$ in the \emph{resonant case} due to $\mu$-distortion for $\bar m_a=2.66\times 10^{-6}-4.88\times 10^{-5}$ eV. In (a) the exclusion plot for COBE \cite{Fixsen:1996nj} upper limit on $\mu$ is shown and in (b) the sensitivity region of PIXIE/PRISM \cite{Kogut:2011xw}, based on the expected sensitivity on the $\mu$ parameter is shown. In both figures the region above the solid line corresponds to the KSVZ axion model ($|C_{a\gamma}|\simeq 4$) and the region above dot dashed line correspond to the DFSZ axion model ($|C_{a\gamma}|\simeq 1.49$).}
\label{fig:Fig1a}
\end{figure*}

In the presence of $\mu$-distortion, we can expand the photon occupation number for, $\mu\ll 1$, in power series and using the fact that leakage of photons is due to oscillations into axions, we get the following relation between $P_a$ and $\mu$
\begin{equation}\label{axion-mu}
P_a=\mu\,\frac{e^{x}}{e^{x}-1}.
\end{equation}
Using Eqs. \eqref{axion-mu}, \eqref{res-probab} and Eq. \eqref{res-temp} we get the following relation between the magnetic field strength and the axion mass
\begin{equation}\label{B-ma}
B_\textrm{nG}=\frac{0.22}{\bar m_a\,C_{a\gamma}}\,\left(\frac{\mu\, e^{x}}{x(e^{x}-1)}\right)^{1/2}.
\end{equation}
We can see that Eq. \eqref{B-ma} depends on the photon energy $x$ and tighter bound on $B_\textrm{nG}$ or $\bar m_a$ are obtained for higher values of $x$. Indeed, using for example the energy range explored by COBE/FIRAS \cite{Fixsen:1996nj}, $1.2\leq x\leq 11.3$ we get a tighter limit on $B_\textrm{nG}$ at $x=11.3$
\begin{equation}\label{final-eq}
B_\textrm{nG}=6.76\times 10^{-2}\,\frac{\sqrt{\mu}}{\bar m_a\,C_{a\gamma}}.
\end{equation}
Eq. \eqref{final-eq} is our main result, which connect three unknown parameters $\bar m_a, B_\textrm{nG}$ and $C_{a\gamma}$, with the $\mu$ parameter which is determined by experiment. We may notice, that for values of $\mu$ given by COBE \cite{Fixsen:1996nj} and PIXIE/PRISM \cite{Kogut:2011xw} we have that the bound given by Eq. \eqref{B-limit} is indeed well satisfied.  We emphasize that our results in the resonant case (see Fig. \ref{fig:Fig1a}), obtained by using Eq. \eqref{final-eq}, perfectly agree with the numerical solution of the quantum kinetic equation, Eq. \eqref{dens-ope}, in the steady state approximation. 


Concluding, our main results are shown in Fig. \ref{fig:Fig1a} when we present the exclusion and sensitivity limits on the magnetic field strength vs. axion mass in the resonant case. In Fig. \ref{fig:Fig1} the exclusion region in the case of COBE is respectively shown for the KSVZ and DFSZ axion model. In Fig. \ref{fig:Fig2} the sensitivity region of future space mission PIXIE is shown. If PIXIE will detect any spectral distortion in the CMB spectrum, that would be a potential signal of photon to axion oscillation. 

In general is not possible to give definite limits on $B$ and $m_a$ since none of them is known exactly and moreover only limits (in the case of COBE, upper limit) on $\mu$ parameter exist which relates both. Nevertheless, we can outline important conclusions considering the upper limits of all of them. We can base our arguments by simply focusing on Eq. \eqref{final-eq}. Firstly, based on the limit on $\mu$ from COBE we can limit the axion mass, if we know the limit on $B$. For instance, in the case of the KSVZ axion model and a homogeneous magnetic field with strength $B\lesssim 3.2$ nG we obtain from Eq. \eqref{final-eq} that $m_a\lesssim 4.8\times 10^{-5}$ eV. The limit on magnetic field strength is by a factor 1.2 stronger than that found for an uniform and anisotropic magnetic field in Ref. \cite{Barrow:1997mj} and is by a factor 3.2 weaker than that found in Ref. \cite{Blasi:1999hu}, from the Faraday rotation of the Layman $\alpha$-forest. It is interesting to note that the upper limit $B\lesssim 3.2$ nG is very close to the limit found in Ref. \cite{Giovannini:2009zq}  ($B\lesssim 3.1-3.2$ nG) from the CMB temperature cross correlation spectra, TT and TE of WMAP 5 yr for the case of stochastic magnetic field with comoving coherence length scale $\lambda_B\simeq 1$ Mpc and field spectral index $n_B\simeq 1.6$ (blue magnetic field spectrum). 

For the DFSZ axion model the upper limit, for an uniform magnetic field is $B\lesssim 9$ nG which is by a factor 2.5 weaker than the KSVZ axion model for the same axion mass, see Fig. \ref{fig:Fig1}. This upper limit on the magnetic field strength for the DFSZ axion model, would produce larger temperature anisotropy with respect to the observed one, and makes the DFSZ axion model disfavoured with respect to the KSVZ axion model.  PIXIE/PRISM are more sensitive than COBE and in principle can better confine the axion parameter space with respect to COBE, see Fig. \ref{fig:Fig2}  limits with respect to it. In particular, in the case of detection of spectral distortions, and assuming that the strength of the magnetic field is close to its canonical value, $B\simeq 1$ nG (for an uniform magnetic  field with coherence length of the Hubble horizon), it would be an extremely important signature of axions in the mass range, $\bar m_a\simeq 2\,\mu$eV-$3\, \mu$eV. 

The ADMX collaboration \cite{ADMX} excluded all axion models of being dark matter in the mass region $3.3\, \mu$eV-$3.5\, \mu$eV.  This mass range lies in the axion mass range considered in this paper, see Eq. \eqref{mass-range}. Thus, it would be possible to use the ADMX limits on the axion mass to constrain the magnetic field strength. For example, considering $m_a\simeq 3.5\, \mu$eV, we find the magnetic field strength to be (in the case of COBE), $B\simeq 46$ nG for the KSVZ axion model and $B\simeq 130$ nG for the DFSZ axion model. In the case of PIXIE/PRISM we would have $B\simeq 1$ nG for the KSVZ axion model and $B\simeq 3$ nG for the DFSZ axion model. However, knowing  the upper and/or the lower limit for the axion mass, it allows us to constrain only the magnetic field strength. In this case, the above limits are valid for an uniform  magnetic field with a coherence length of at least comparable with the horizon scale during the $\mu$ epoch,  $\lambda_B^\mu\sim H^{-1}(z_\mu)$ or $\lambda_B^\mu(z_\mu)\sim$ 3.8 pc (or $\lambda_B^\mu\sim 1.3$ Mpc at present) where the redshift corresponding to the resonant axion mass $\bar m_a\simeq 3.5\,\mu$ eV during the $\mu$ epoch  is $z_\mu\simeq 3.44\times 10^5$, see Eq. \eqref{res-temp}.

The derived limits for an uniform magnetic field with coherence length comparable with the horizon scale, are in general stronger than those found from the temperature anisotropy \cite{Barrow:1997mj} and slightly weaker than those found from the Faraday rotation \cite{Blasi:1999hu}, at smaller coherence length scales. Indeed, at the coherence length scale $\lambda_B\simeq 1$ Mpc, the Faraday rotation of the Lyman $\alpha$-forest gives $B\lesssim 10$ nG \cite{Blasi:1999hu} which is by a factor 5.3 stronger than the limit found for the KSVZ axion model and by a factor 14.1 stronger than the DFSZ axion model (using ADMX limit on the axion mass and the COBE limit on the $\mu$ parameter). The limits on the axion mass found here, in general, are of the same order of magnitude, with the limits found by the misalignment mechanism, see Ref. \cite{Sikivie:2006ni} and \cite{pdg}. Indeed, the upper limit $m_a\lesssim 4.8\times 10^{-5}$ eV for $B\lesssim 3.2$ nG is very close to that found in Ref. \cite{DiValentino:2014zna}, $m_a\lesssim 76\,\mu$ eV-82$\,\mu$ eV for CDM axions. According to Ref. \cite{DiValentino:2014zna} an axion within this mass range would explain all dark matter contents in the universe without requiring other candidates. In our case an axion in the mass range $m_a\lesssim 7.6\,\mu$ eV-8.2$\,\mu$ eV would make non resonant oscillation into CMB photons during the $\mu$-epoch. If the misalignment mechanism limits are used instead of the ADMX limit, for the axion mass range (non resonant oscillation) coincident with those in Ref. \cite{DiValentino:2014zna}, the strength of homogeneous magnetic field at $\lambda_B\simeq 1$ Mpc would be between $1.4\times 10^3$ nG - $1.6\times 10^3$ nG depending on the nonresonant axion mass. These limits are weaker than those found from the Faraday rotation of the Lyman $\alpha$-forest and are comparable with the limits found from a homogeneous universe, see Ref. \cite{Blasi:1999hu}.

\end{document}